\documentclass{article}

\usepackage{arxiv}

\usepackage[utf8]{inputenc} 
\usepackage[T1]{fontenc}    
\usepackage{hyperref}       
\usepackage{url}            
\usepackage{booktabs}       
\usepackage{amsfonts}       
\usepackage{nicefrac}       
\usepackage{microtype}      
\usepackage{lipsum}		
\usepackage{graphicx}
\usepackage{physics, amsmath}
\usepackage{doi}
\usepackage[title]{appendix}

\usepackage[nottoc]{tocbibind}
\usepackage{xcolor}

\title{Fish Sounds: towards the evaluation of marine acoustic biodiversity through data-driven audio source separation} 

\author{
\href{https://orcid.org/0000-0000-0000-0000}{\includegraphics[scale=0.06]{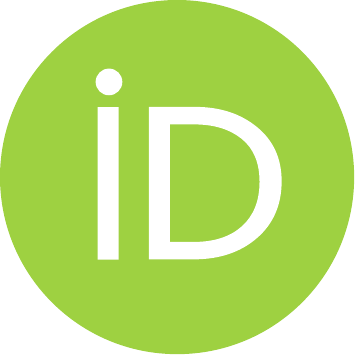}\hspace{1mm}Michele Mancusi} \\
	\textit{Sapienza University of Rome}\\
	Rome, Italy \\
	\texttt{mancusi@di.uniroma1.it} \\
	\And
    Nicola Zonca \\
	\textit{Studio Arki}\\
	Bologna, Italy \\
	\texttt{nicola@studioarki.com} \\
	\And
	\href{https://orcid.org/0000-0000-0000-0000}{\includegraphics[scale=0.06]{orcid.pdf}\hspace{1mm}Emanuele Rodolà} \\
	\textit{Sapienza University of Rome}\\
	Rome, Italy \\
	\texttt{rodola@di.uniroma1.it} \\
	
	\And
	\href{https://orcid.org/0000-0003-1358-0828}{\includegraphics[scale=0.06]{orcid.pdf}\hspace{1mm}Silvia Zuffi} \\
	\textit{IMATI-CNR}\\
	\textit{Consiglio Nazionale delle Ricerche}\\
	Milan, Italy \\
	\texttt{silvia@mi.imati.cnr.it} \\
}

\renewcommand{\shorttitle}{\textit{arXiv} Template}

\begin{document}
\maketitle

\begin{abstract}
The marine ecosystem 
is changing at an alarming rate, exhibiting biodiversity loss and the migration of tropical species to temperate basins. Monitoring the underwater environments and their inhabitants
is of fundamental importance to understand the evolution of these systems and implement safeguard policies. However, assessing and tracking biodiversity is often a complex task, especially in large and uncontrolled environments, such as the oceans. One of the most popular
and effective methods for monitoring marine biodiversity
is passive acoustics monitoring (PAM), which employs hydrophones to capture underwater sound.
Many aquatic animals produce sounds characteristic of their own species; these signals travel efficiently underwater and can be detected even at great distances. Furthermore, modern technologies are becoming more and more convenient and precise, allowing for very accurate and careful data acquisition.
To date, audio captured with PAM devices is frequently manually processed by  marine biologists and interpreted with traditional signal processing techniques for the detection of animal vocalizations. This is a challenging task, as PAM recordings are often over long periods of time. Moreover, one of the causes of biodiversity loss is sound pollution; in data obtained from regions with loud anthropic noise, it is hard to separate the artificial from the fish sound manually.
Nowadays, machine learning and, in particular, deep learning represents the state of the art for processing audio signals. 
Specifically, sound separation networks are able to identify and separate human voices and musical instruments. In this work, we show that the same techniques can be successfully used to automatically extract fish vocalizations in PAM recordings, opening up the possibility for biodiversity monitoring at a large scale.
\end{abstract}


\section{Introduction}
The ocean covers 71\% of the Earth's surface and represents the natural habitat of numerous marine species. The biodiversity present in this environment is impressive, and keeping track of the activity and quantity of all these species is essential for monitoring the marine ecosystem. In fact, today more than ever, the environmental issue is of crucial importance, and the ocean, like the whole planet Earth, is facing drastic and dramatic changes due to human activity, some of which are overfishing, disease and ocean warming \cite{harvell1999}. These changes, in addition to damaging the marine ecosystem, mainly affect the species that live there, therefore, monitoring biodiversity is of vital importance to understand the trend of the abundance of marine fauna, identify the most vulnerable areas and take action to safeguard endangered species \cite{balvanera2006}. However, this monitoring is not simple at all because many of the methods used on the Earth's surface for tracking species, such as photos and videos, are often not so effective in the marine environment due to the inability to access many places and due to poor light and visibility conditions. Furthermore, much data that can be collected on physical quantities, such as temperature, salinity and pressure, are not indicative of the biodiversity in a given place. Therefore, there is a need for a tool capable of overcoming these obstacles and providing an accurate analysis of the biodiversity of the marine habitat. Instead of exploiting optical signals, sound signals can be used to monitor biodiversity \cite{pijanowski2006}. In fact, the acoustic environment faithfully reflects the characteristics of the fauna and its behavior present in that place \cite{southworth1969}. 

One of the most popular
and effective methods for monitoring marine biodiversity
is passive acoustics monitoring (PAM), which employs hydrophones to capture underwater sound.
Many aquatic animals produce sounds characteristic of their own species, while modern technologies are becoming more and more convenient and precise, allowing for very accurate and careful data acquisition.
Acoustic indices were initially used to assess biodiversity from PAM recordings \cite{sueur2008rapid}. These indices are used to estimate richness, amplitude, heterogeneity, and evenness of an acoustic environment. Some of these indices are, for example, the acoustic entropy index (\textit{H}), which indicates how much the amplitude of a signal is uniform in time and frequency; or the acoustic complexity index (\textit{ACI}) which takes into account the variation of a signal in different frequency bins over time, and then averages over the entire frequency range \cite{PIERETTI2011868}. 
While easy to apply, a drawback of acoustic indices is that they are not learned from data, and they are not, therefore, discriminative for animal sounds with respect to sounds with similar patterns but of a different origin.  
When the goal is the detection of fish vocalization, typically the PAM audio signal is visualized as a spectrogram and visually inspected from an expert. 
This task exploits the fact that fishes vocalize in a relatively small range of low frequencies and often produce repetitive sounds.

In this work, we aim at automating the process of fish sound detection by means of a network that can separate fish sounds from the sea background, fully automatically. Specifically, we employ recent advances in sound separation for human speech and music to the problem of separating fish sound from PAM recordings.
Machine learning techniques, and deep learning, in particular, require a large amount of data. In supervised learning, data need to be annotated to provide ground truth information for training neural networks. Data annotation usually entails the manual identification of the attributes one wants to automatically recover. Obtaining annotated data for the task of sound separation, given a mixed signal, is clearly too challenging, and a preferable approach is to generate the training data by mixing individual sounds.
This approach has been largely exploited for speech, music and anthropic sound. But while for human speech and music there is an abundance of data 
, this is not the case for fish vocalizations. 
At present, to our knowledge, there are no datasets available that include many examples of fish voices.
Therefore, we collected a dataset of fish vocalizations from the Web, obtaining, in the majority of the cases, a single sound example for each species. This limits the possibility of applying AI techniques to automatically classify the fish species from sound, as several recordings for each species would be necessary. However, fish sounds from different species exhibit similarities and have characteristics that make it possible to train a network that can separate the sound produced by fishes from the background sound.  
We created a sound separation dataset by randomly overlapping fish vocalizations with sea backgrounds that we recorded in different locations on the Greek island of Nisyros. We use this dataset to train two recent and popular architectures for sound separation: Conv-TasNet (which we will call TasNet) and Demucs. We quantitatively evaluate the networks' performance on a synthetic test set. We quantitatively show performance on a few examples of recordings performed in Marsa Alam, Egypt. Our quantitative evaluation shows that the sound emitted from fishes can be successfully recovered from recordings with simulated noisy background. 
To our knowledge, this is the first work that applies modern sound separation techniques to PAM.

\section{Related Work}
The assessment of marine biodiversity through acoustic techniques is evolving rapidly and although several methods are used, for the moment none of these is considered the ideal tool for investigating the marine environment and its diversity. Some principal methods are discussed below.
\subsection{Classical approaches}
Spectrograms are helpful tools to study the trend over time of the amplitude of the different frequencies present in an audio signal.
In this way, through the Fourier transform, it is possible to visualize a 1D signal in a 2D image. By the study of spectrograms, it is possible to identify 
the presence of some marine species through the 
vocalizations they emit. For example, it is possible to observe whether the sound emitted is rhythmic or more smooth and harmonious. 
The analysis of spectrograms as images
allows the observation of approximate spectral patterns, but cannot accurately identify the detailed structure of sounds modulation occurring in the oceans. Therefore, automatic systems based on pattern recognition in images
are not sufficient for detecting fish vocalizations, and spectrograms needs often to be checked manually.

As for long-term recordings, lasting months or years, it is unthinkable to do a manual analysis of all the recorded material, both in terms of the time used and the people involved. Long-term record structure analysis is critical to understanding the factors influencing changes in marine biodiversity. 
One of the techniques used to analyze spectrograms is clustering \cite{rijsbergen1979}. Assuming that the data has underlying patterns, this method allows grouping elements with similar characteristics. Hence, large amounts of audio data can be grouped into a few audio clusters and these can be exploited to assess biodiversity by measuring acoustic metrics. Unfortunately, this type of analysis can easily fail when non-biological sources contaminate the collected data \cite{lin2018comparison}. However, it is necessary to individually analyze the sources that are part of each cluster to understand the key elements that contribute to marine biodiversity.

\subsection{Learning-based approaches}
Sound travels faster in liquids than in gases, and low frequencies can travel for miles. Therefore, the marine acoustic scenario can often be very complicated due to the interference of the sounds that propagate in the ocean. Several studies \cite{innami2012nmf, lin2018listening, gillespie2009pamguard, zhang2017blind, xie2016adaptive, jiang2019whistle, luo2019convolutional, stowell2019automatic} have been carried out to trace, recognize, and isolate the biological sound sources present in nature and the use of machine learning has been fundamental to obtaining significant results. In particular, in \cite{luo2019convolutional, stowell2019automatic}, the authors propose using the most advanced deep learning techniques for the detection task of odontocete echolocation and bird sounds, respectively. For the task of source separation of marine biological signals, more straightforward and older automatic methods are used, respectively non-negative matrix factorization (NMF) and the Sawada algorithm~\cite{innami2012nmf, zhang2017blind}. Finally, in the context of music and speech separation, deep learning has contributed significantly to achieving impressive performance \cite{luo2019conv, defossez2019music}. 


\section{Method}

The problem of separating audio sources consists of breaking down a mixture of signals $y(t) \in \mathbb{R}^{T}$ into its $n$ components $c_1(t), \ldots, c_n(t) \in \mathbb{R}^{T}$, where,
\begin{align}
y(t) = \sum_{i=1}^n c_i(t).
\label{eqn:prob}
\end{align}
The mixture is represented as a vector in the waveform domain. In our case, we have $n=2$ sources: fish and background.

In order to perform the sound separation, we employ the two aforementioned networks, TasNet and Demucs. These two types of networks are trained in a supervised manner, and while both have an encoder-decoder structure and act directly on the audio waveform, they are fundamentally different. In fact, TasNet learns a mask to apply to the mixture to filter the desired signal, whereas Demucs learns to directly synthesize the required signals without using any filtering.
We train both networks with supervision using the same synthetic generated data, as explained in the following sections. 


\subsection{TasNet}
\begin{figure}
\includegraphics[width=1.0\textwidth]{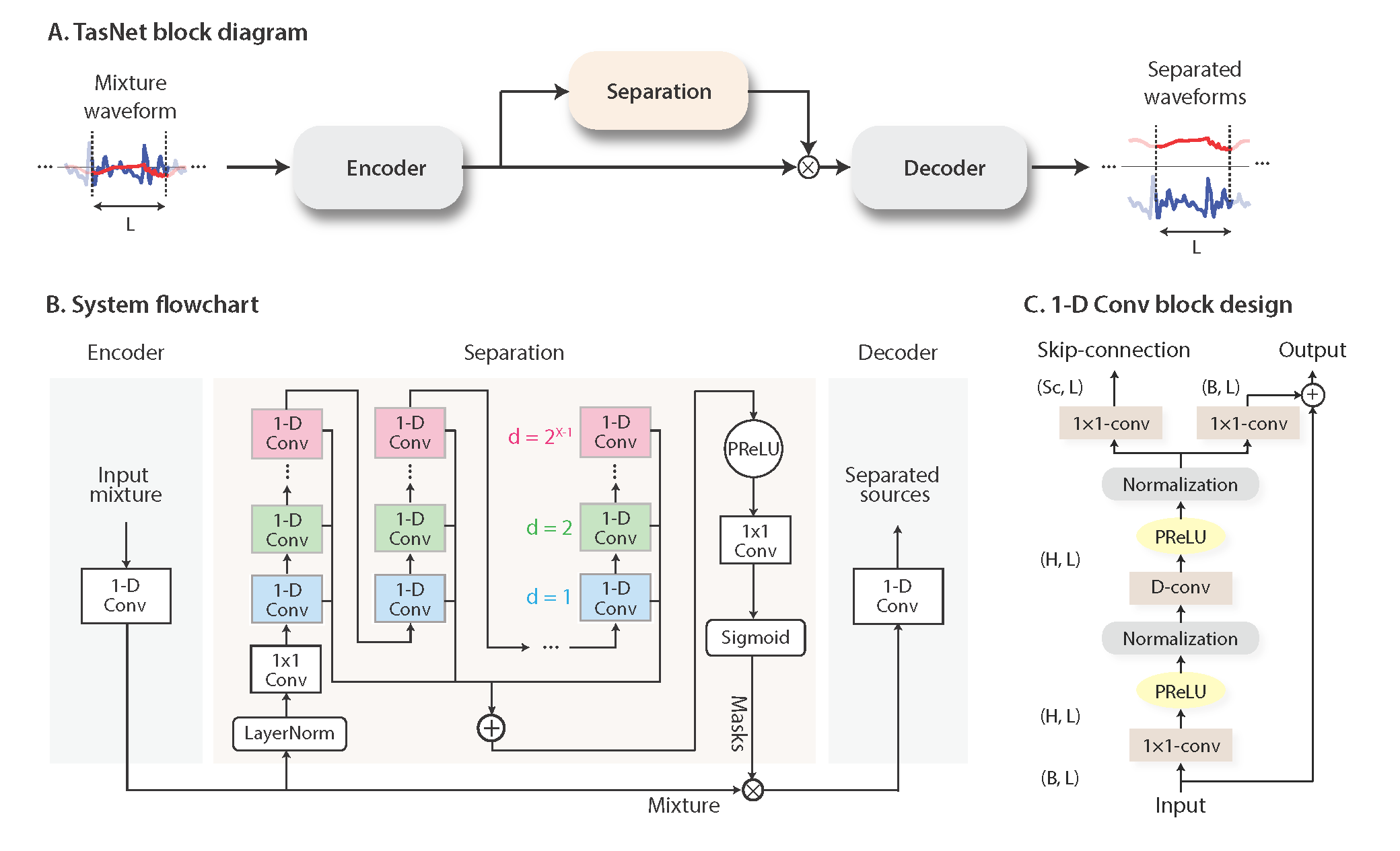}
\caption{(A): TasNet block diagram. A slice of the mixture is mapped into a high-dimensional latent space, then a separation module estimates a mask for each source and in the end a decoder transforms the masked encoded features into signals in the waveform domain. (B): System flowchart. The encoder consists of a 1D convolutional module that maps the mixture into the features space. A temporal convolutional network (TCN) calculates the mask vectors, and the decoder reconstructs the separated signals by a 1D transposed convolution operation. In the separation module, different dilation factors in each 1D Conv block are highlighted with different colors. (C): 1D convolutional block design. Each block is made of a $1\times 1$ convolution ($1\times 1$-conv), followed by a depthwise convolution ($D$-conv) \cite{chollet2017xception}. Between these two modules, there are a nonlinear activation function (PReLU) and a normalization layer. At the end of the block there are two more $1\times 1$-conv blocks: one is for the skip connections and the other for the residual path.
This figure is taken from \cite{luo2019conv}.}
\label{fig:tasnet}
\end{figure}

TasNet is a convolutional audio separation model in the time domain, and it is composed by an encoder, a separation module and a decoder, as shown in figure~\ref{fig:tasnet} (A). The encoder has the role of transforming small overlapping fragments of the mixture into feature vectors in an intermediate latent space. Using this representation, the separation module calculates a mask for each source. Each of these, multiplied by the respective intermediate representation of the mixture, generates the latent features of the relative source. Finally, the decoder converts each latent representation into a time-domain waveform, thus obtaining the desired signals. In figure ~\ref{fig:tasnet} (B) we show the entire system flowchart.

\subsubsection{Encoder}
Initially, the input mixture is divided into $N$ overlapping parts $\vectorbold{x}_i \in \mathbb{R}^{L}$, where $i = 1,\dots,{N}$, each of length $L$. Each $\vectorbold{x}_i$ is transformed by the encoder into the corresponding vector in the latent domain $\vectorbold{z}_i \in \mathbb{R}^{M}$ through a 1D convolution operation (which is formally expressed by a matrix multiplication) followed by a ReLU activation function $\mathcal{G}(\cdot)$:
\begin{align}
\vectorbold{z}_i = \mathcal{G}(\vectorbold{x}_i\vectorbold{S})\,,
\label{eqn:enc}
\end{align}
where $\vectorbold{S}$ is a ${L\times M}$ matrix containing the convolution coefficients.

\subsubsection{Separation module}
The actual separation of each fragment of the mixture occurs in the separation module, in which $n$ mask vectors $\vectorbold{m}_{i} \in \mathbb{R}^{M}$ are estimated, where $i=1, \ldots, n$ and $n$ is the number of signals to be separated. Each of these vectors, being masks, must necessarily be $\vectorbold{m}_{i} \in [0, 1]$. The vector representation in the latent space $\vectorbold{b}_i \in \mathbb{R}^{M}$ of each signal is calculated by multiplying the relative mask $\vectorbold{m}_i$ by the mixture $\vectorbold{z}_i$,
\begin{gather}
\vectorbold{b}_i = \vectorbold{z}_i \odot \vectorbold{m}_i
\label{eqn:mask}
\end{gather}
where $\odot$ denotes element-wise multiplication. 
This module 
is a temporal convolutional network (TCN) \cite{lea2016temporal}, which is fully convolutional and consists of stacked 1D dilated convolutional blocks with increasing dilation factors. These factors make it possible to gradually capture increasingly broad contexts, thus exploiting long-range dependencies within the signal. In figure~\ref{fig:tasnet} (C) we show the 1D convolutional block design.

\subsubsection{Decoder}
The reconstruction of each source is computed by the decoder. The latter takes as input $\vectorbold{z}_i$ and returns a vector $\hat{\vectorbold{x}}_i$ in the waveform domain by applying a 1D transposed convolution operation,
\begin{align}
\hat{\vectorbold{x}}_i = \vectorbold{z}_i\vectorbold{T}
\label{eqn:dec}
\end{align}
where $\hat{\vectorbold{x}}_i \in \mathbb{R}^{L}$ is the reconstruction of $\vectorbold{x}_i$ and $\vectorbold{T}$ is a ${M\times L}$ matrix of convolution weights.

\subsection{Demucs}
Demucs is an autoencoder model made of a convolutional encoder and a convolutional decoder linked with skip U-Net connections and a 2-layers bidirectional LSTM. 
The size of the latent space is $C_B=6$.
\subsubsection{Encoder}
\begin{figure}
\includegraphics[width=1.0\textwidth]{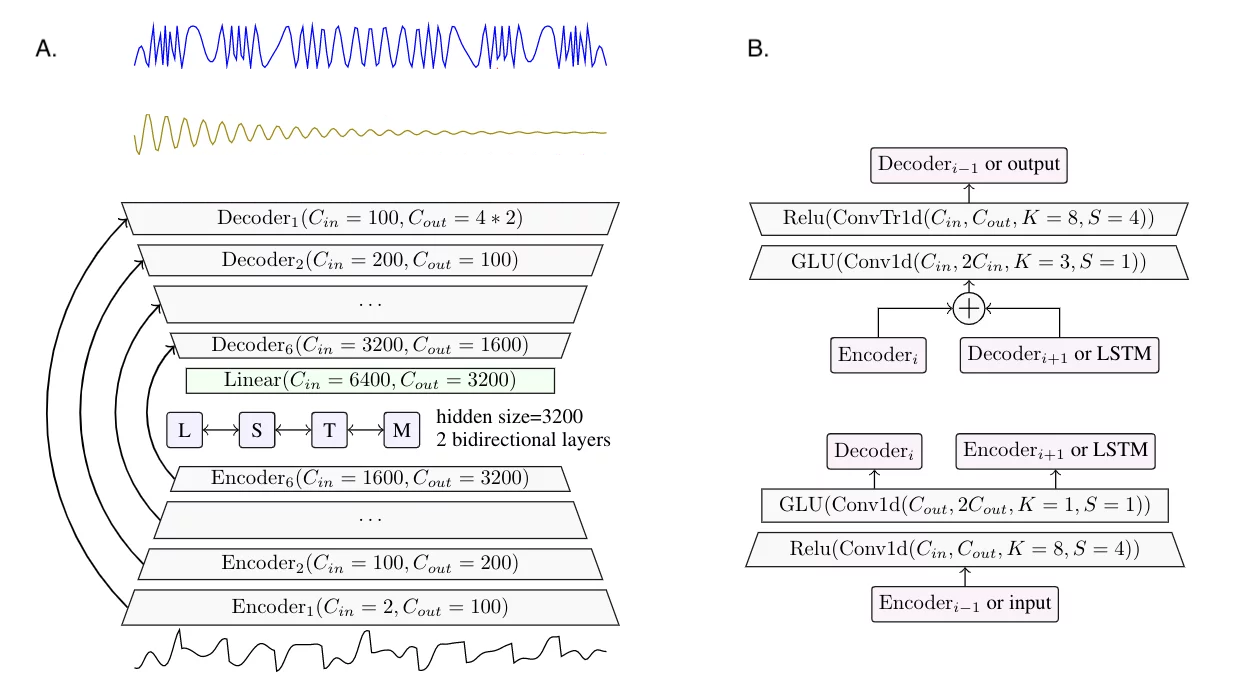}
\caption{(A): Demucs model with the input mixture and the two output sources, all in the waveform domain. (B): Encoder/decoder block architecture. In each encoder block, there is a convolution with kernel size $K=8$ (to have dependencies with adjacent time steps) and stride $S=4$ followed by a ReLU activation function. The result is given as input to another convolution with kernel size $K=1$ and stride $S=1$, in order to increase the expressivity of the network with little additional computation. In the end, a gated linear unit (GLU) activation function \cite{dauphin2017language} is applied. The decoder block is constructed in reverse order with respect to the encoder, and it consists of convolution with kernel size $K=3$ and stride $S=1$, followed by a GLU and then a transposed convolution with kernel size $K=8$ and stride $S=4$, followed by a ReLU. This figure is taken from \cite{defossez2019music}.}
\label{fig:demucs}
\end{figure}

As illustrated in figure~\ref{fig:demucs}, the encoder consist of $B=6$ stacked convolutional layers and the number of output channels $C_{i}$ in each layer equals the number of input channels $C_{i+1}$ in the next layer. From the second layer onwards, the output channels are twice the number of input channels. All these stacked layers have the task of compressing the information in order to obtain a compact representation of the training data. The input channels in the first layer are $C_{0}=2$ and the output channels are $C_{0}=100$. The output channels in the last layer are $C_{B}=3200$, which is the hidden size of the LSTM.
\subsubsection{Decoder}
Since LSTM outputs a tensor with $2C_{B}$ channels, a linear layer is needed to reduce the number of channels to $C_{B}$. The decoder is built essentially like the encoder, but with the convolutional layers put in reverse order and transposed convolutions instead of the regular convolutions. The decoder has the task of expanding the dimensions of the compressed vectors in the latent space to regain vectors with sizes equal to those of the input space.
The last layer returns tensors with $N\cdot C_{0}$ channels, synthesizing the $N$ sources present, initially, in the input mixture.
\subsubsection{U-network}
In this architecture, the encoder layers are connected to the decoder layers with the same index through skip connections, as happens in the Wave-U-Net \cite{jansson2017singing}. The objective of these connections is to connect the various decoder layers with those of the encoders to transfer information directly from ones to the others in such a way as to facilitate reconstruction. Compared to Wave-U-Net, Demucs skip connections use transposed convolutions instead of linear interpolations, since they require less memory and computational time.

\section{Experiments}
\label{sec:experiments}
In this work, we employ two types of data: registrations of fish vocalizations and sea recordings.

\subsection{Fish Vocalization Data}
\begin{figure}
\includegraphics[width=1.0\textwidth]{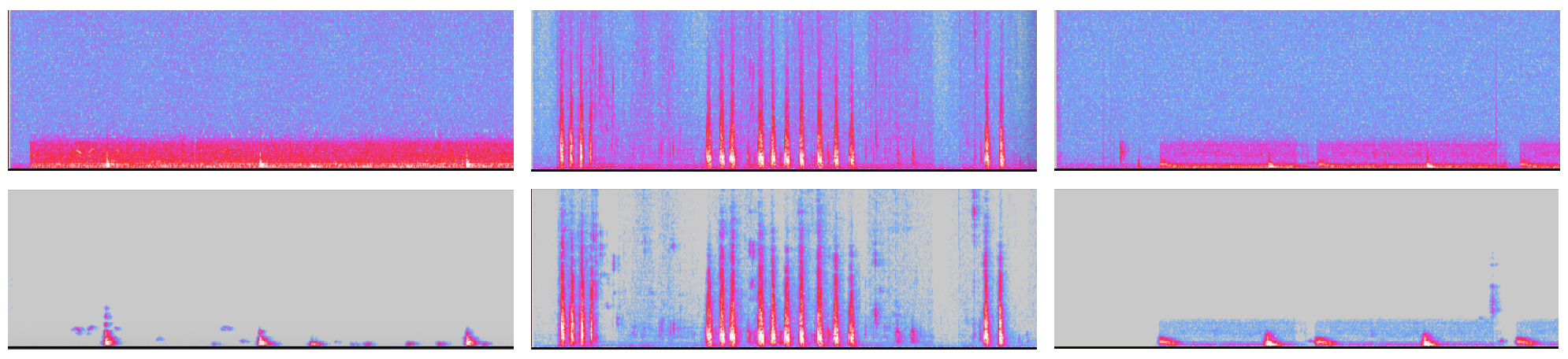}
\caption{Fish vocalization before (top) and after (bottom) noise removal and normalization.} 
\label{fig:voices}
\end{figure}

We collected 191 audio files corresponding to the vocalization of 143 different species. Most of the recordings were downloaded from FishBase\footnote{\url{www.fishbase.org}}. 
The collected data often exhibit unnatural noise, since in many cases the recordings are performed in fish tanks. 
In order to create a dataset that can be used to synthesize realistic audio data, 
we preprocessed for noise removal, and normalized for a peak amplitude of $-1$ dB. For this purpose, we used the open source software Audacity. Figure \ref{fig:voices} illustrates examples before and after preprocessing.
We employ the fish vocalizations for the online creation of training samples by combining them with recorded sea backgrounds and for creating a synthetic testset with ground-truth separated signals.

\subsection{Sea Recordings}
We performed sea recordings at the Greek island of Nisyros.
Recordings were performed in October 2019, April 2021, August 2021 and October 2021 at different sites around the island, both near the coast and in the open sea. Data was captured with an Aquarian Scientific AS-1 hydrophone. 
Sea recordings at Nisyros are used as sea backgrounds.
In addition, we collected a sound video dataset at Marsa Alam using an action camera. The audio channel from this data is used for a qualitative evaluation.

\subsection{Networks Training}
We train the network on a synthetic training set obtained by combining the vocalization dataset (foreground) with sea recordings (background).
We considered $11$ sea recordings for creating backgrounds.
The first 5 files were captured with sample rate of $192K$ and were converted to $44K$. Recordings with length greater than $3$ minutes were divided in multiple files of smaller duration ($1-2$ minutes) and constitute a dataset of background chunks. We have a total of $133$ files for background.
A couple of recordings were manually filtered for removing fish sounds. This was not necessary for the other recordings, where fish sounds were harder to find.

Audio data is loaded from the network as a set of audio chunks of length $44160$, with a $0.25$ overlap. The synthetic data used for training is created as follows.
We load the vocalization data as a set of samples, where each sample is a chunk.
During training, for each sample $i$ required from the network, we define the two audio sources $s_0$ and $s_1$:

\begin{align*}
    s_0 = k_f \alpha_f x_f \\
    s_1 = (1+k_b) x_b,
\end{align*}

where $x_f$ is the sample with index $i$, and $x_b$ is a random background chunk; $k_f$ and $k_b$ are two random coefficients sampled from a uniform distribution, while $\alpha_f$ is a fixed attenuation factor for the fish audio. In this way, at each epoch, every fish sample is combined differently with a random background. We set $\alpha_f = 0.1$.
We use the $80\%$ of the fish vocalization data for training. We trained both the networks with a learning rate of $1e-4$ and a number of epochs equal to $200$.

\subsection{Evaluation}
For the quantitative evaluation, we generated a set of synthetic inputs using the $20\%$ fish vocalizations that were not used for training, combining these sounds at random with background chunks out of the training distribution. 
For the qualitative evaluation, we trained the network using the whole vocalization dataset, and show results 
in Figure \ref{fig:voicesma} and \ref{fig:voicesma2}.

We apply the trained TasNet and Demucs networks to the synthetic testset. We quantify the sound separation performance, computing an SDR (Source to
Distortion Ratio) score \cite{vincent2006performance} between recovered and ground truth fish and background audio. In particular, the reconstruction $\hat{s}_i$ of a source $s_{target}$ can be thought of as consisting of four components:
\begin{align*}
    \hat{s}_i = s_{target} + e_{interf} + e_{artif} + e_{noise}
\end{align*}

where $e_{interf}$, $e_{artif}$ and $e_{noise}$ are respectively error terms for interference, artifacts and noise \cite{vincent2006performance}. Using these terms, it is possible to calculate different metrics expressed in $dB$, including the SDR:

\begin{align*}
    \mathrm{SDR} := 10\log_{10}{\left(\frac{\norm {s_{target}}^2}{\norm{e_{interf} + e_{artif} + e_{noise}}^2}\right)} \,.
\end{align*}

In general, SDR is considered an excellent metric to assess sound quality (the higher, the better). 
Table \ref{tab:results} reports our results. 

\begin{table}[h!]
	\caption{Quantitative evaluation on the synthetic testset. In this table, it can be seen how TasNet performs better than Demucs.}
	\centering
	\begin{tabular}{lll}
		\toprule
		\multicolumn{2}{c}{TasNet}                   \\
		\cmidrule(r){1-3}
		Metric     & Channel     & Value \\
		\midrule
		SDR & Fish  & $\textbf{10.59}$     \\
		SDR     & Background & $\textbf{17.60}$      \\
		
		\bottomrule
	\end{tabular}
		\begin{tabular}{lll}
		\toprule
		\multicolumn{2}{c}{Demucs}                   \\
		\cmidrule(r){1-3}
		Metric     & Channel     & Value \\
		\midrule
		SDR & Fish  & $-5.96$     \\
		SDR     & Background & $2.65$      \\
		
		\bottomrule
	\end{tabular}
	\label{tab:results}
\end{table}
\newpage

\section{Conclusions}
In this work, we have seen how deep learning techniques used in signal processing can be effectively applied on marine data for source separation. In particular, we note how the Tasnet network performs significantly better than Demucs. The former reaches an SDR score equal to $10.59$ on the separation of the sound of the fish and $17.60$ on the background, while the latter obtains just $2.65$ of SDR  on the background and even a negative score on fish, equal to $-5.96$ of SDR. Although the separations produced by Demucs appear to be perceptibly better, from spectrograms in the appendix, it can be seen how they show artifacts; in particular, vertical lines are introduced that are repeated periodically, while in Tasnet, this behavior is not present. Furthermore, it is possible to notice how, on the synthetic data, in correspondence with the sounds of the fish, Demucs generates fictitious frequencies that are not present in the TasNet separations. This is probably due to the fact that Demucs is a network that does not separate the signal by filtering it, but by directly synthesizing the requested source, not performing well with the data of our dataset. Instead, Tasnet, a more classical network that filters the desired signal from the mixture, appears to be more robust and performs better with the data in our possession. In general, in deep learning, neural networks need many data to obtain good results, and it is very encouraging that with our relatively scarce data, certainly less than those used to train these networks in the field of speech and music separation, such SDR scores have been achieved. We hope these results will pave the way for new methods of studying the marine environment and contribute to developing new automatic PAM techniques for monitoring marine biodiversity and, possibly, accurately tracking fauna in the oceans. 

\section*{Acknowledgments}
MM and ER are supported by the ERC grant no. 802554 (SPECGEO).

\bibliography{references}
\newpage
\begin{appendices}
\section{Spectrograms examples}
\vspace{-1.em}
\begin{figure}[h!]
\includegraphics[width=1\textwidth]{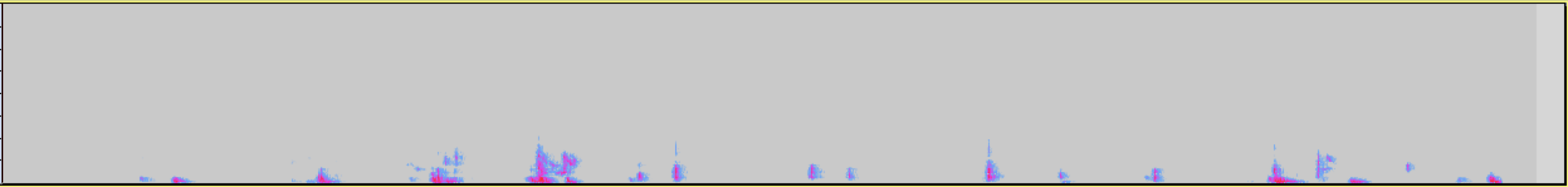} 
Sea background + ground truth fish (network input):\\
\includegraphics[width=1\textwidth]{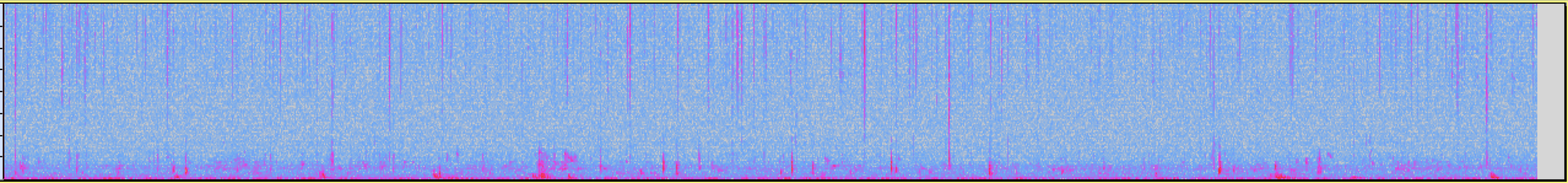}
Estimated fish (TasNet result):\\
\includegraphics[width=1\textwidth]{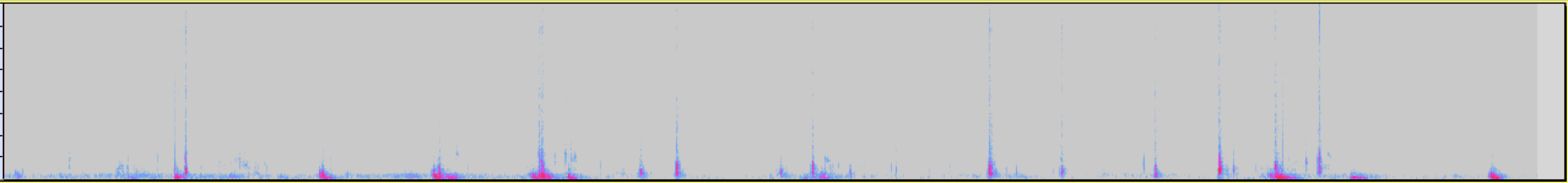}
Estimated fish (Demucs result):\\
\includegraphics[width=1\textwidth]{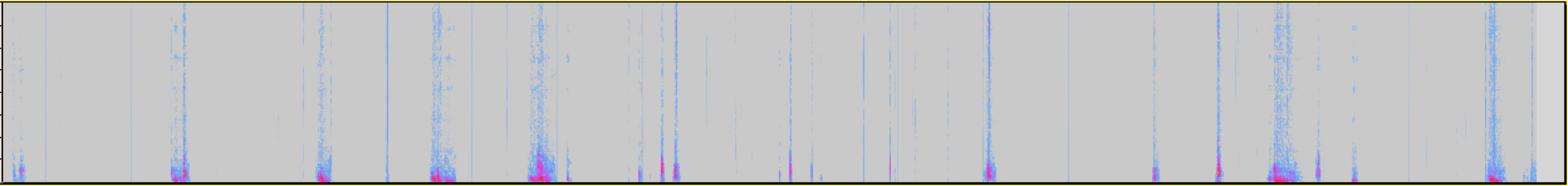}
\vspace{-2.em}
\caption{Synthetic testset example. From top: fish vocalization; overlap with sea background; TasNet fish separation; Demucs fish separation.} 
\label{fig:stest01}
\end{figure}

\begin{figure}[h!]
\includegraphics[width=1\textwidth]{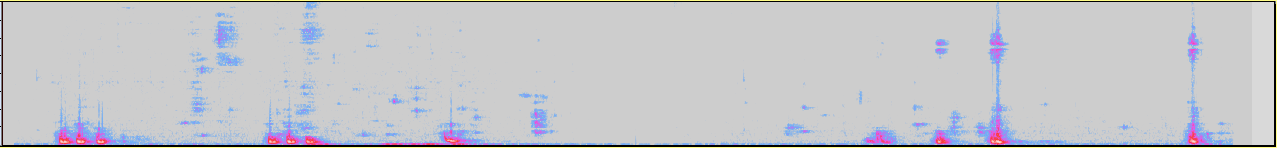}
Sea background + ground truth fish (network input):\\
\includegraphics[width=1\textwidth]{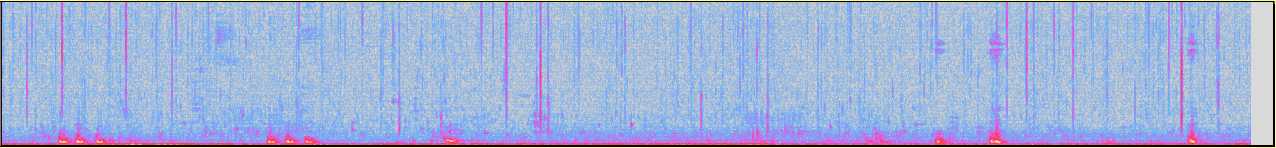}
Estimated fish (TasNet result):\\
\includegraphics[width=1\textwidth]{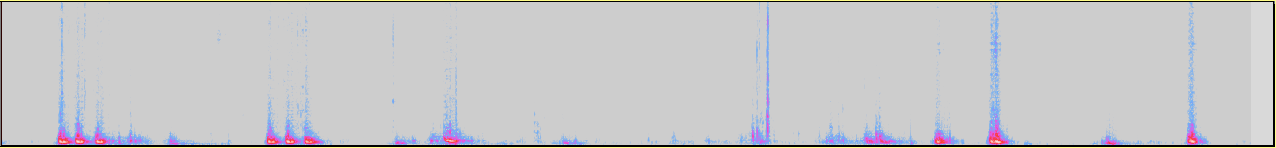}
Estimated fish (Demucs result):\\
\includegraphics[width=1\textwidth]{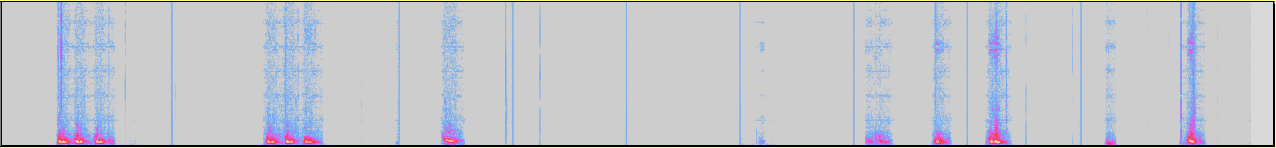}
\vspace{-2.em}
\caption{Synthetic testset example. From top: fish vocalization; overlap with sea background; TasNet fish separation; Demucs fish separation.} 
\label{fig:stest02}
\end{figure}

\begin{figure}
Ground truth fish:\\
\includegraphics[width=1.0\textwidth]{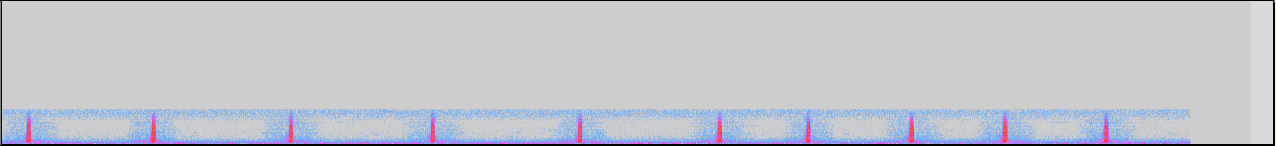}
Sea background + ground truth fish (network input):\\
\includegraphics[width=1.0\textwidth]{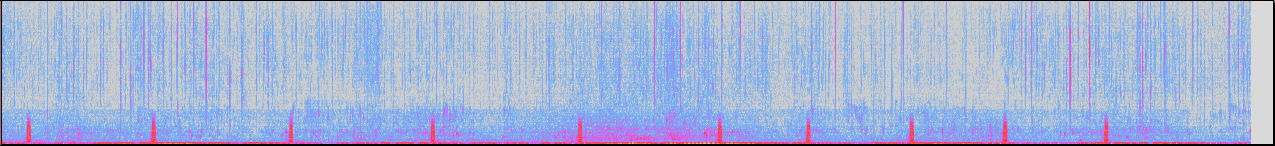}
Estimated fish (TasNet result):\\
\includegraphics[width=1.0\textwidth]{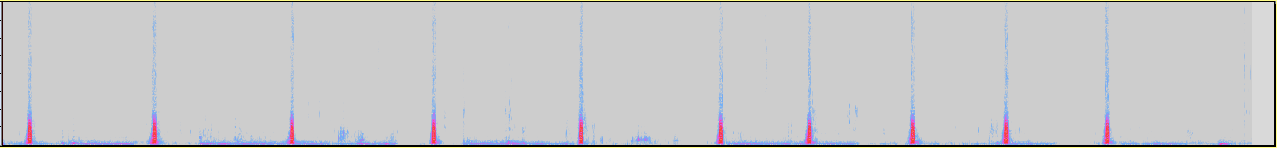}
Estimated fish (Demucs result):\\
\includegraphics[width=1.0\textwidth]{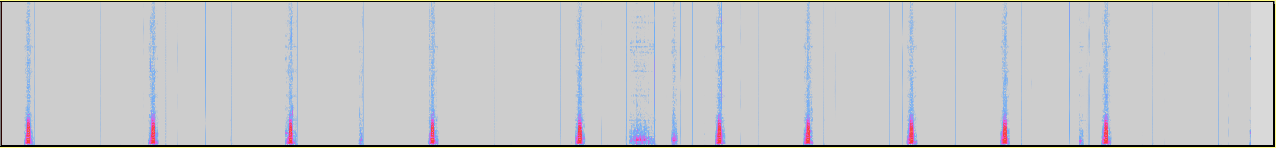}
\vspace{-.5em}
\caption{Synthetic testset example. From top: fish vocalization; overlap with sea background; TasNet fish separation; Demucs fish separation.} 
\label{fig:stest03}
\end{figure}


\begin{figure}
Network input:\\
\includegraphics[width=1.0\textwidth]{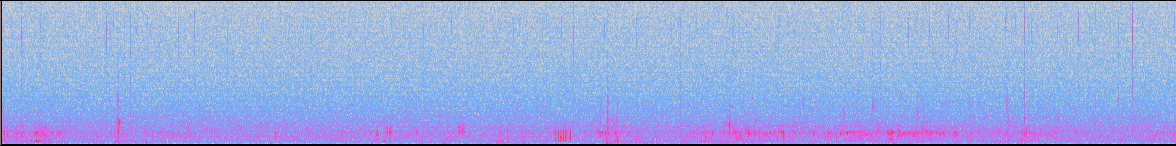}
TasNet result:\\
\includegraphics[width=1.0\textwidth]{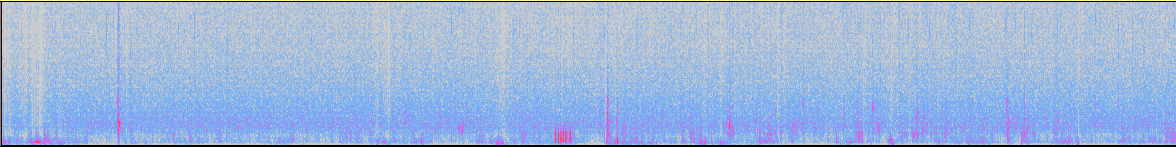}
Demucs result:\\
\includegraphics[width=1.0\textwidth]{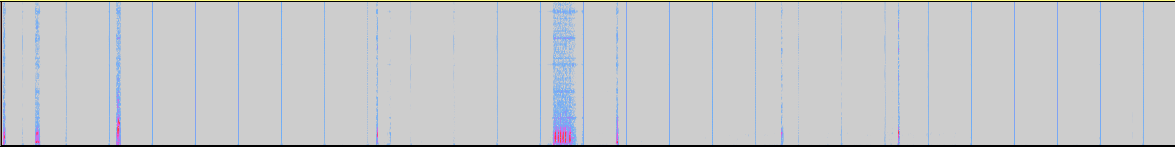}

\caption{Sea recording at Marsa Alam (top, 21 sec); fish output channel (middle) obtained with TasNet; (bottom) with Demucs.}
\label{fig:voicesma}
\end{figure}

\begin{figure}
Network input:\\
\includegraphics[width=1.0\textwidth]{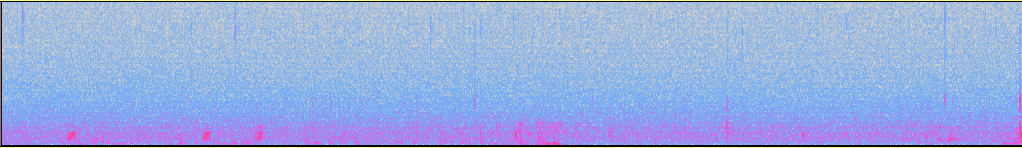}
TasNet result:\\
\includegraphics[width=1.0\textwidth]{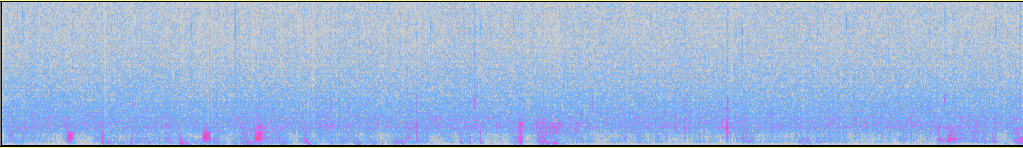}
Demucs result:\\
\includegraphics[width=1.0\textwidth]{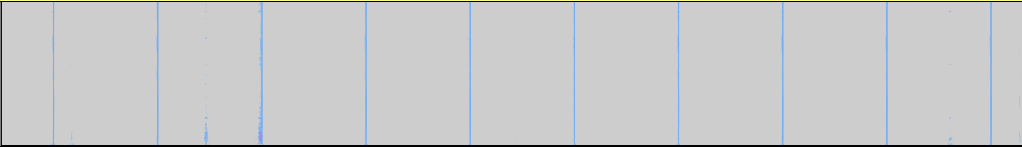}

\vspace{-.5em}
\caption{Sea recording at Marsa Alam (top, 9 sec); fish output channel (middle) obtained with TasNet; (bottom) with Demucs.}
\label{fig:voicesma2}
\end{figure}

\begin{figure}
Network input:\\
\includegraphics[width=1.0\textwidth]{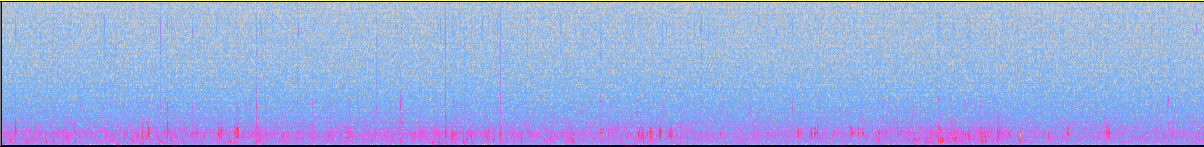}
TasNet result:\\
\includegraphics[width=1.0\textwidth]{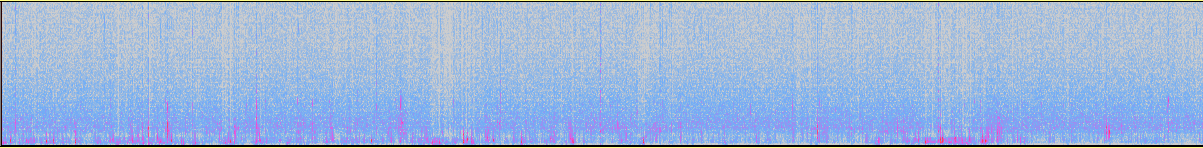}
Demucs result:\\
\includegraphics[width=1.0\textwidth]{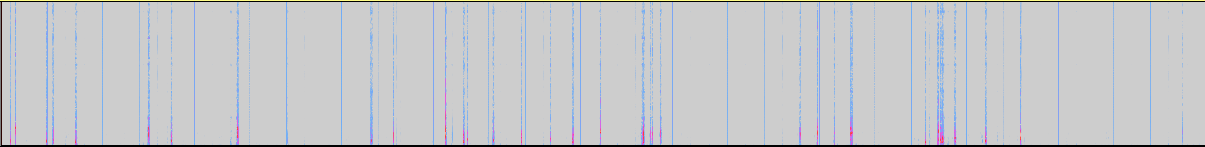}

\vspace{-.5em}
\caption{Sea recording at Marsa Alam (top, 60 sec); fish output channel (middle) obtained with TasNet; (bottom) with Demucs.}
\label{fig:voicesma3}
\end{figure}

\end{appendices}

\end{document}